\documentstyle[twocolumn,prl,aps,epsfig]{revtex}
\begin{document}
\twocolumn[\hsize\textwidth\columnwidth\hsize\csname @twocolumnfalse\endcsname

\draft \title{Asymmetric magnetic interference patterns  in 
0-$\pi$ Josephson junctions.}
\author{D.F. Agterberg$^1$ and  M. Sigrist$^{1,2}$}
\address{$^1$ Theoretische Physik, ETH-
H\"{o}nggerberg, 8093 Z\"{u}rich, Switzerland \\
$^2$ Yukawa Institute for Theoretical Physics, Kyoto University, Kyoto 
606-01, Japan}
\date{\today}
\maketitle
\begin{abstract}
We examine the magnetic interference patterns 
of Josephson junctions with a region of 0- and of $ \pi $-phase
shift. Such junctions have recently been realized as $c$-axis YBCO-Pb
junctions with a single twin boundary in YBCO.
We show that in general  the junction generates self-fields which
introduces an  asymmetry in the critical current under reversal
of the magnetic field. Numerical calculations of these asymmetries 
indicate that they account well for unexplained features observed
in single twin boundary junctions.
\end{abstract}

\pacs{74.20.Mn,74.25.Bt}
\vskip2pc]
\narrowtext

Josephson tunneling experiments on the high temperature superconducting
compound YBa$_2$Cu$_3$O$_7$ (YBCO) have played an important role in
establishing the predominantly 
$d-$wave symmetry of the superconducting order parameter
\cite{wol93,bra94,mat95,igu94,wol95,tsu94,kir95,mil95}.  
However, YBCO 
has orthorhombic symmetry which implies that the order parameter cannot be
purely $d$-wave and must contain an additional $s$-wave contribution
\cite{sun95,sig96,wal96,odo95,li93}. This has 
significant consequences for $c$-axis Josephson tunneling
experiments between YBCO and Pb (a standard $s$-wave superconductor).
In such junctions the $d$-wave component is forbidden by symmetry to 
contribute to the lowest order Josephson
current so that the observed current \cite{sun95}
is due solely to the $s$-wave component of the YBCO. There is a
clear difference between untwinned and twinned YBCO samples in the
observed current. The
current is considerably suppressed for the latter compared to
that of the former. This can be understood if we assume
that the $d$-wave component is essentially uniform while the $s$-wave
component changes sign at each twin boundary, {\it i.e.} a twinned sample
yields a $c$-axis junction with alternating 0- and $\pi$-phase shift
regions. This alternation of sign (or phase) gives rise to destructive 
interference effects for the total Josephson current in heavily
twinned samples \cite{sig96}.

To gain more insight into the $c$-axis 
Josephson tunneling results Kouznetsov {\it et. al.} \cite{kou97} 
have built $c$-axis YBCO - Pb junctions that contain
a single twin boundary and have measured its magnetic interference pattern.
These elegant experiments show that the critical current
as a function of the applied magnetic field, $I_c({\bf H}_e)$,
behaves qualitatively like a junction separated into two regions, one
with 0- and the other with $ \pi $-phase shift (such a model naturally arises due to $\pi$ phase change 
of the $s$-wave component as the twin boundary is crossed). 
A noteworthy unexplained feature of the 
data in Ref.~\cite{kou97} is an asymmetry between $I_c({\bf H}_e)$
and $I_c(-{\bf H}_e)$ for fields in the junction plane that are neither along the
twin boundary nor orthogonal to it. 
In the short junction limit (for which the junction 
dimensions are much smaller than the Josephson penetration 
depth) such an asymmetry implies that time reversal symmetry 
is broken in the junction when there are no applied currents or 
magnetic fields
\cite{agt97}. For example such an asymmetry will occur if there is local
broken time reversal symmetry due to a phase difference of $\pi/2$ between 
the $s$ and $d$ components at the twin boundary (far from the twin boundary this 
phase difference
must be $0$ or $\pi$) \cite{agt97,kou97}.   

\begin{figure}
\epsfxsize=80mm
\epsffile{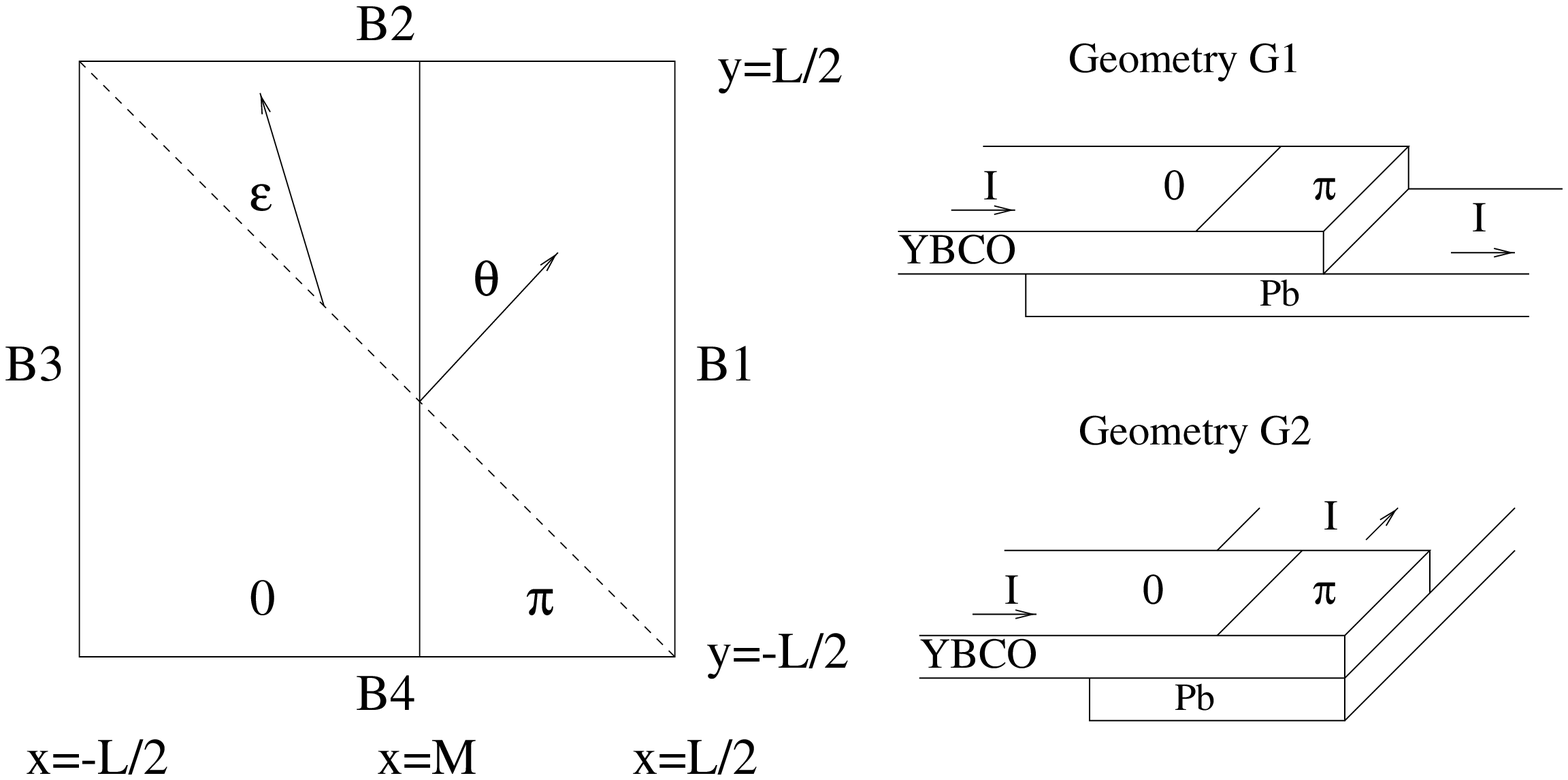}
\vglue 0.1 cm
\caption[*]{Junction geometries considered in this article.
The left figure shows the junction plane while the right
figures 
depict the two geometries considered in this article.}
\label{fig1}
\end{figure}

In this article we present another origin 
for such an asymmetry in junctions that are 
time reversal invariant. 
In particular, such asymmetries arise whenever
parity symmetry is broken in these junctions. 
This parity breaking can occur in two ways:
either the junction is not invariant  
under a parity 
transformation about the center of the junction (intrinsic
parity breaking) or the
current running from the leads into the junction does not obey 
$I|_{\partial S}=I|_{-\partial S}$ where $\partial S$ is the boundary
of the junction ({\it e.g} the current does not 
enter and leave
 a rectangular  junction
through opposite edges) (extrinsic parity breaking). 
In the context of the $0-\pi$ junction model for a single
twin boundary the junction has intrinsic parity symmetry 
whenever the twin boundary lies in the center of the junction
(called a symmetric junction after Kirtley {\it et. al.} 
\cite{kir97}).
A numerical study for a symmetric $0-\pi$ junction 
with extrinsic broken parity symmetry is shown to account
 well for the
 experimentally 
observed magnetic diffraction patterns.
Numerical studies of asymmetric $0-\pi$ junctions  
with extrinsic parity conserving boundary conditions
also reveals 
asymmetries that should be observable in experiments
like those
of Ref.~\cite{kou97}.
This discussion is also
of relevance to corner junction flux modulation experiments
\cite{wol95,mil95,xu95}. 

We consider here junctions with the two geometries shown in Fig.1. 
The geometry $G2$ breaks extrinsic parity symmetry  
while $G1$ does not.
The coordinates perpendicular and parallel to the twin
boundary are called $ x $ and $ y $, respectively. 
The order parameter
of the Pb has standard $s$-wave symmetry, $ \Psi_{Pb} = | \Psi_{Pb} |
e^{i \varphi} $. For YBCO there are two components $ (\Psi_d , \Psi_s )
= e^{i \varphi'} (|\Psi_d|, e^{i \alpha} | \Psi_s | ) $ where $ \Psi_d
$ and $ \Psi_s $ correspond to the $d$- and $s$-wave components,
respectively. The relative phase $ \alpha $ is fixed by the bulk free
energy of YBCO and is 0 for one type of twin domain and  $\pi $ for
the other \cite{sig96,wal96}. The intrinsic phase shift of the
junction is determined by 
$ \alpha $. Note that the choice of $ \alpha $ equal to 0 or $ \pi $ is a 
matter of convention (freedom of gauge in the two
superconductors). However, once it is fixed somewhere in the 
junction it is determined everywhere. 

The Josephson phase, the local phase difference $ \phi = \varphi - \varphi'
$ over the junction follows the relation 
$ {\bf H} ({\bf x})\times \hat{z} = (\Phi_0 /
2\pi\tilde{d}) ( { \bf \nabla} \phi ) $ 
while $ {\bf H}$ satisfies the Maxwell equation 
$ \nabla \times  {\bf H}  = 4 \pi {\bf j}
/ c $ ($ \Phi_0 = hc/2e $) \cite{tin89}. 
For the lowest order Josephson coupling
(between $ \Psi_s $ and $ \Psi_{Pb} $) we can express $ j_z $ as
$ j_c({\bf x}) \sin [ \phi + \alpha({\bf x}) ] $ ($ j_c({\bf x}) > 0 $). 
By restricting $\alpha({\bf x})$ to take on the values $0$ or $\pi$ 
and allowing $j_c({\bf x})$ to  take on any value this relation for
$j_z$ describes any time reversal invariant junction. 
We therefore use this form for more general considerations and
then specify $j_c({\bf x})$ and $\alpha({\bf x})$ to describe the
$0-\pi$ model for the single twin boundary junction for 
numerical results.  
Together these
relations lead to the Sine-Gordon equation describing the spatial
dependence of $ \phi $ throughout the bulk of the junction
\cite{sig96,wal96,tin89}.
\begin{equation}
\nabla^2\phi
=\lambda_J^{-2}({\bf x})\sin[\phi+\alpha({\bf x})],
\end{equation}
where
$\lambda_J({\bf x} )
=[\Phi_0/2\pi\tilde{d}j_c({\bf x})]^{1/2}$ 
with the boundary conditions
\begin{equation}
\bar{\lambda}_J{\bf n}\cdot \nabla\phi |_{\partial S}
= {\bf n}\cdot ({\bf  h}\times \hat{z})|_{\partial S}
\end{equation} 
where ${\bf n}$ is the outward normal to the boundary $\partial S$.
The variable ${\bf h}$ is a function of the reduced variables 
${\bf h}^e=\frac{2ed\bar{\lambda}_J}{\hbar c} {\bf H_e}$ (${\bf H}_e$ 
is the applied
external field) and 
$i=\frac{I}{2\bar{\lambda}_J L \bar{j}_c}$ where $\bar{j}_c$ is
the average value of $j_c({\bf x})$ and $\bar{\lambda}_J$ is
determined from $\bar{j}_c$. For the geometries shown in
Fig 1 Ampere's Law implies  \cite{owe67} that we have for G1
(extrinsic parity conserving boundary conditions)
\begin{equation}
{\bf n}\cdot[{\bf h}\times \hat{z}]|_{\partial S}= 
\left \{ \matrix{h^e_y+i&\partial S=B1\cr
-h^e_x&\partial S=B2\cr
-h^e_y+i&\partial S=B3\cr
h^e_x&\partial S=B4. }\right .
\end{equation} and for G2 (extrinsic parity violating boundary conditions)
\begin{equation}
{\bf n}\cdot[{\bf h}\times \hat{z}]|_{\partial S}= 
\left \{ \matrix{h^e_y&\partial S=B1\cr
-h^e_x+i&\partial S=B2\cr
-h^e_y+i&\partial S=B3\cr
h^e_x&\partial S=B4. }\right . 
\end{equation}
The Sine-Gordon equation and the boundary conditions can be 
recast into a variational form with the functional
\begin{equation} \begin{array}{l}
F=\bar{\lambda}_J \int_{S}d^2x \left[\left(\nabla \phi\right)^2/2
-\lambda_J^{-2}({\bf x})\cos(\phi+\alpha({\bf x}))
\right ] \\ \\ +\int_{\partial S} dx \phi[{\bf n}\cdot({\bf h}\times \hat{z})].
\end{array} \label{free}
\end{equation}
The positive and negative critical currents of the junction 
are  defined as the maximal
positive and negative supercurrents, respectively, that can pass through the
junction. They satisfy the 
following symmetry relations based on the structure of $ F  $:
\begin{eqnarray}
I_c[{\bf H}_e,\alpha({\bf x}),\lambda_J({\bf x})]=&-I_c[-{\bf
  H}_e,-\alpha({\bf x}),\lambda_J({\bf x })] \label{sym1}\\
I_c[{\bf H}_e,\alpha({\bf x}),\lambda_J({\bf x})]=&I_c[{\bf H}_e,\alpha({\bf
  x})+\alpha_0,\lambda_J({\bf x})]
\label{sym2}
\end{eqnarray}
where $\alpha_0$ is a constant phase shift. 
Eqs. \ref{sym1} and \ref{sym2} follow from the time reversal and gauge
invariance of the junction respectively. 
For geometry G1, due to the extrinsic parity symmetry, 
the following relation also holds
\begin{equation}
I_c[{\bf H}_e,\alpha({\bf x}),\lambda_J({\bf x})]=I_c[-{\bf H}_e,\alpha(-{\bf
  x}),\lambda_J(-{\bf x})]
\label{sym3}
\end{equation} more specifically this  is a result of 
the invariance of $F$ in Eq.~(\ref{free}) under the transformation
${\bf h}^e\rightarrow -{\bf h}^e$, $\alpha({\bf x})
\rightarrow \alpha(-{\bf x})$,
$\lambda({\bf x})\rightarrow\lambda(-{\bf x})$, and $\phi({\bf x})\rightarrow
\phi(-{\bf x})$ (a
product of time reversal and parity symmetries). 
No such relation holds for geometry 
$G2$ which has broken extrinsic parity (this is true for 
any geometry that exhibits extrinsic parity breaking). 
If the junction is time reversal invariant then the
phase $\alpha({\bf x})$ is restricted to be $0$ or $\pi$.
For such a 
junction $I_c({\bf H}_e)=I_c(-{\bf H}_e)$ is required if 
the boundaries have extrinsic parity symmetry, 
if $j_c({\bf x})=j_c(-{\bf x})$, and if $\alpha({\bf x})=\alpha(-{\bf x})$ 
or if $\alpha({\bf x})=\alpha(-{\bf x})+\pi$.
Consequently the observation
of $I_c({\bf H}_e)\ne I_c(-{\bf H}_e)$ in junctions that are time
reversal invariant is due to broken parity symmetry in the
junction.
In the following we will consider the $0-\pi$ model for
the single twin boundary junction where $ \lambda_J ({\bf x}) =
\lambda_J $ and $ \alpha({\bf x}) = \pi \Theta(x-M) $ where  $ \Theta $ is
the step function and $ - L/2 < M < + L/2 $ ($ M $ is the position of the
twin boundary according to Fig.1).

Consider the geometry G1. The one dimensional model that 
results
when the magnetic field is applied 
along the twin boundary has been previously studied
by several groups (e.g. see Ref. \cite{kir97,xu95}). Using
the relations (\ref{sym2}) and (\ref{sym3}) with $\alpha_0=\pi$ 
it is easy to see that
$I_c({\bf H}_e)=I_c(-{\bf H}_e)$ if the twin boundary lies 
in the center of the
junction (a symmetric junction \cite{kir97}). For other positions of the twin boundary
there is no symmetry constraint to enforce  $I_c({\bf H}_e)\ne I_c(-{\bf H}_e)$,
however it can be shown for these asymmetric $0-\pi$  junctions that
\begin{equation}
I_c(h_x^e,h_y^e)=I_c(-h_x^e,h_y^e).
\label{sym4}
\end{equation} 
Numerical analysis shows 
$I_c({\bf H}_e)\ne I_c(-{\bf H}_e)$ when $h_y^e\ne0$ for asymmetric $0-\pi$ junctions 
(see below).
As pointed out earlier \cite{xu95} the change of $ \alpha $ from 0 to
$ \pi $ introduces a spontaneous flux line corresponding to a ``$ \pi
$ vortex'', i.e. vortex of half the conventional phase winding and a
flux $ \Phi_0/2$, if the junction is much longer than the screening
length $ \lambda_J $ \cite{millis94}. Even for junctions with a length comparable to $
\lambda_J $ such self
fields appear. However, in this case the screening is imperfect and
new effects due to the position of the twin boundary arise. 
For an asymmetric junction the screening is more effective 
for the longer of the $\alpha=0$ or $\alpha=\pi$ regions 
leading to $I_c({\bf H}_e)\ne I_c(-{\bf H}_e)$. 
Note that the time reversal
invariance of asymmetric $0-\pi$ junctions requires $I_c({\bf H}_e)=-I_c(-{\bf H}_e)$
so that
experiments in which the positive and negative critical currents are
averaged (such as in Ref. \cite{mil95}) will not reveal this 
asymmetry. A recent careful numerical
study of this model when the magnetic field is along the
twin boundary by Kirtley {\it et. al.} \cite{kir97} only considered 
one sign of the applied
field and consequently did not uncover this asymmetry.
\begin{figure}
\epsfxsize=85mm
\epsffile{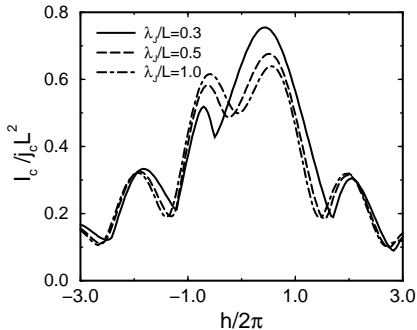}
\caption[*]{$I_c(H_e)/j_cL^2$ for geometry G1,$M/L=0.25$,
 and $\lambda_J/L=$0.3, 0.5, and $1.0$ from top to bottom
respectively. $I_c(H_e)$ represents the positive
 critical
current and the negative critical current is given by $-I_c(-H_e)$.}
\label{fig2}
\end{figure}
To examine the magnitude of the difference between
$I_c({\bf H}_e)$ and $I_c(-{\bf H}_e)$ for asymmetric $0-\pi$ junctions
a numerical calculation is required. Since the two dimensional problem
is numerically intensive a detailed study for the field along the twin
boundary
(for which the problem becomes one dimensional) was done. 
The technique we use is to 
discretize the variable $\phi(x)$ into $N$ variables and to minimize 
the free energy by
a quasi-Newton technique. The critical current was found by increasing
the current until no stable minimum can be found anymore. The criterion
for failure to reach a stable state in $F$ in  Eq. \ref{free} 
is similar to that used by Kirtley {\it et. al.} \cite{kir97} 
and
is given by $\epsilon>10^{-4}$ where 
$(\epsilon/\pi)^2=[\phi_1-\phi_2-(i-h)/\alpha^{1/2}]^2
+[\phi_N-\phi_{N-1}-(i+h)/\alpha^{1/2}]^2+
\sum_{i=2}^{N-1}[\phi_{i+1}+\phi_{i-1}-2\phi_i-
\sin(\phi_i+\theta_i)/\alpha]^2$ where $\alpha=N\lambda_J^2/L^2$.
Our numerical results using $N=100$ for the symmetric $0-\pi$ junction and
 for the $0-0$ 
junction agree well with those found by Kirtley {\it et. al.} \cite{kir97}. 
Fig. 2 shows the results for $M/L=1/4$ and 
for a variety of penetration depths. 
The asymmetry between $I_c(H_e)$ and $I_c(-H_e)$ is clearly visible
and this further implies that zero field does not correspond to 
either a maximum
or a minimum of the magnetic diffraction pattern.

Now consider the geometry G2.
This geometry is the relevant one for the  experiments on single twin boundary
junctions \cite{kou97} and in this
case  there are
no symmetry relations enforcing $I_c({\bf H}_e)=I_c(-{\bf H}_e)$ for
symmetric $0-\pi$ junctions due to extrinsic parity breaking.  
To gain an understanding of  the
effects of this geometry on the $I_c({\bf H}_e)$ pattern it is helpful
to first consider $0-0$ junctions ({\it i.e.} $\lambda_J({\bf x}) =
\lambda_J $ and $\alpha({\bf x})=0$ independent of $ {\bf x} $). Symmetry relations
enforce $I_c({\bf H}_e)=I_c(-{\bf H}_e)$ 
when the field is along the dotted diagonal shown in Fig. 1. These relations
also imply
$I_c(H_e,\epsilon)=I_c(-H_e,-\epsilon)$ where $\epsilon$ is defined in Fig. 1.
 A 2D generalization (for a 30 by 30
system and $L/\lambda_J=5$) of the 1D numerical method presented 
earlier shows  that as expected 
a conventional 
$I_c({\bf H}_e)$ for $\epsilon=0$ arises 
and as $\epsilon$ increases
the central peak tilts so that the maximum $I_c$ moves from $H_e=0$ to positive fields.
The maximum $I_c$ is furthest away from $H_e=0$ for $\epsilon=\pi/2$. 
We have not been able to find any experimental reports of this 
angular dependent $I_c({\bf H}_e)$ pattern for  conventional 
Josephson junctions.
 To some degree
this angular asymmetry still occurs for the $0-\pi$ junction. For the
$0-\pi$ junction the interplay of the presence of the $0-\pi$ phase
shift and the geometry makes a qualitative discussion difficult. In Fig.3
we present numerical results on a 30 by 30 system for $\lambda_J=1/2$.
The numerical results exhibit all of the main features
observed in experiments: the asymmetry is not apparent for the
field along  or perpendicular to the twin boundary but 
is so for other field orientations, the sign of $I_c({\bf H}_e)-I_c(-{\bf H}_e)$
changes when the field passes through the twin boundary, and the magnitude of
the asymmetry is in good agreement with the experimentally 
quoted $\lambda_J/L$ values.
 Note that the asymmetry is large for the 
$\theta=\pi/4$ and small for $\theta=-\pi/4$ (in fact it appears to
vanish between $\theta=-\pi/4$ and $\theta=-\pi/3$)  as was the
case for the $0-0$ junction.  This agreement between experiment and 
theory gives further support
 to the claim that the $s$-wave component in the YBCO superconductor
changes sign as the twin boundary is crossed. These
results are for a 30 by 30 system and studies for the 1D system show
that increasing the system size does not change the qualitative 
behavior of the $I_c({\bf H}_e)$ patterns but does increase the size of  
the critical currents and also moves the minima in the $I_c({\bf H}_e)$
patterns  
towards zero field. This explanation for the experimental results
 has the consequence that 
$I_c({\bf H}_e)=-I_c(-{\bf H}_e)$ as follows from the
time reversal symmetry of the junction. A deviation from this
equality implies that time reversal symmetry is broken in the junction.
 It is also of interest to study 
experimentally the geometry G1 since some qualitative differences in the
$I_c({\bf H}_e)$ patterns between the two geometries 
 are
expected to arise.  For example in geometry G1 the asymmetry should vanish for
 symmetric $0-\pi$ junctions and for asymmetric junctions as $h_x^e$
 is reversed  $I_c({\bf H}_e)-I_c(-{\bf H}_e)$
should be the same due to Eq.~\ref{sym4} (in contrast to changing sign
as it does for geometry G2).
\begin{figure}
\epsfxsize=100mm
\epsffile{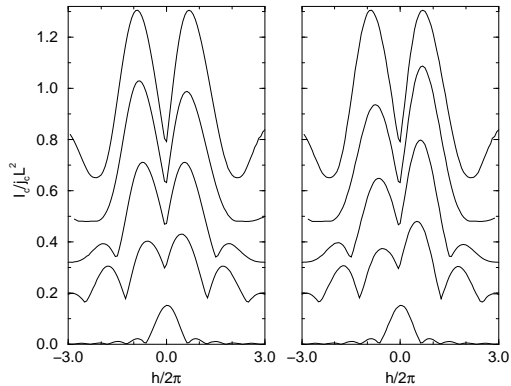}
\caption{$I_c({\bf H}_e)/j_cL^2$ for geometry G2, a symmetric
$0-\pi$ junction, and $\lambda_J/L=0.5$. From top to bottom correspond
to $\theta=0,\pm \pi/6,\pm \pi/4, \pm \pi/3, \pi/2$ where positive (negative) $\theta$ is on the 
right (left) side. Successive plots have been offset by 0.16.}
\label{fig3}
\end{figure}

In conclusion we have shown that parity breaking in 
time reversal invariant Josephson junctions lead to 
an asymmetry in the critical current as the magnetic 
field is reversed. A detailed examination of 
the magnetic diffraction patterns
of $0-\pi$ junctions has shown that self-fields 
account well for unexplained features observed in single
twin boundary junctions.  Experimental tests have
been proposed  to further examine this theory.

We are grateful for the financial support of the Swiss Nationalfonds.
In particular, M.S. was supported by a PROFIL-Fellowship
and D.F.A. by the Zentrum for Theoretische Physik.
D.F.A. acknowledges financial support from the Natural Sciences
and Engineering Research Council of Canada. We also thank T.M. Rice 
and A. Schnidrig for many helpful discussions and B. Chen,
A. Katz  and J. Clarke for useful correspondence on their experiments.

\end{document}